\documentstyle[11pt,aaspp4]{article}

\journalid{337}{}
\articleid{11}{14}

\def\spose#1{\hbox to 0pt{#1\hss}}
\def\lta{\mathrel{\spose{\lower 3pt\hbox{$\mathchar"218$}}
     \raise 2.0pt\hbox{$\mathchar"13C$}}}
\def\gta{\mathrel{\spose{\lower 3pt\hbox{$\mathchar"218$}}
     \raise 2.0pt\hbox{$\mathchar"13E$}}}
\def\n{\noindent}

\def\be{\begin{equation}}
\def\ee{\end{equation}}
\def\msun{M_{\odot}}

\def\mdot{\dot M}

\def\mdote{\dot M_{\rm E}}

\begin{document}

\title {The Spectral States of Black Hole X-ray Binary Sources}
\author{Xingming Chen\altaffilmark{1} and Ronald E. Taam\altaffilmark{2}}

\n \altaffilmark{1}{Department of Astronomy \&
Astrophysics, G{\"o}teborg University and Chalmers University of Technology, 
412 96 G{\"o}teborg, Sweden}

\n \altaffilmark{2}{Department of Physics \& Astronomy, Northwestern 
University, Evanston, IL 60208}

\begin{abstract}
A framework for the interpretation of the spectral states of black hole 
X-ray transients based on the diversity of accretion disk models is 
introduced.
Depending on the mass accretion rate, it is proposed that the accretion disk
is described by one or a combination of the following structures: optically 
thick disk, advection-dominated disk, corona-disk, and non-steady state 
disk.  In particular, it is suggested that the very high, high, 
low, and off states are characterized by mass accretion rates of decreasing
magnitude.  The very high state corresponds to 
mass accretion rates near the Eddington limit in which an optically thin non 
steady inner region is surrounded by an optically thick structure. In 
the high state, the inner region is optically thin and advection-dominated or 
optically thick.  The low hard state is interpreted in terms of a disk-corona 
system and the off state in terms of an optically thin disk dominated by 
advective energy transport into the black hole.  The possible observational
consequences of such a paradigm are discussed. 
\end{abstract}


\section{INTRODUCTION}

One of the most intriguing properties of black hole candidate X-ray sources
(BHCs) is their distinct spectral states.
The first known BHC, Cyg X-1, was discovered by Bowyer et al. (1965)
and was recognized as probably containing a black hole 
by Webster \& Murdin (1972) and Bolton (1972).
Cyg X-1 has an estimated mass in excess of $\sim 7\,\msun$ and exhibits 
high and low states as defined by the X-ray flux in 1--10\,keV band. 
The luminosity difference in this energy band 
between these states is approximately an order of 
magnitude. In the high state, the spectrum consists of two components, 
namely, a relatively stable soft blackbody component and a weak, highly 
variable hard power-law component. On the other hand, in the low state, the 
intensity shows rapid chaotic 
temporal fluctuations with fractional rms amplitudes of several times $10\%$
and which do not depend strongly on photon energy. The spectrum 
is exceptionally hard and can be described as a power-law with a photon 
index $\alpha_N$ less than 2 in the energy band between $\sim 10$ keV 
and a few 100 keV (Liang \& Nolan 1984).
Other persistent BHCs include LMC X-3 and LMC X-1.
Another type of BHC is the soft X-ray transient among which includes
A 0620--00 (V616 Mon), Nova Muscae (GS/GRS 1124--68), and GS 2000+25 
(Nova Vul). A recent review of the observational data obtained from 
these sources has been given, for example, by
Tanaka \& Lewin (1995).

The spectral and time-dependent behavior of BHCs contain
valuable information about the underlying physics of the accretion process.
It is generally believed that the soft blackbody spectral component
from these sources emanates from an optically thick cool accretion disk
(Shakura \& Sunyaev 1973). 
The properties of the low hard state and the power-law component in the high
state imply the existence of optically thin, hot matter. 
However, the structure of the hot optically thin flow is unclear. 
Specifically, several types of models have been suggested including
the optically thin two-temperature disk 
(Shapiro, Lightman, \& Eardley 1976), and various
corona-disk models (Liang \& Price 1977; Ionson \& Kuperus 1984; 
Haardt \& Maraschi 1993; Svensson \& Zdziarski 1994).  Although the
dynamics and stability properties of the corona-disk models have
not been well established, 
it is well-known that the optically thin disk models are 
catastrophically thermally unstable (Piran 1978).

BHC X-ray transients may provide important clues for a model description of 
the different spectral states since they exhibit a wide intensity range in 
which they can be studied.
In the case of Nova Muscae 91, it exhibited all the spectral states
during its decay after the burst in January 1991 (
Kitamoto et al. 1992, Ebisawa et al. 1994).
Since the X-ray nova outburst is very probably due to 
mass flow modulations in the disk due to a thermal 
limit-cycle instability in the accretion disk and/or a mass transfer 
induced instability in the companion star 
(see a recent review by Lasota 1996), it is conceivable
that the decay of the light curve reflects the decrease of the mass flow 
rate in the disk. Near the peak of the burst, the soft X-ray blackbody 
component and the hard X-ray, power law component, are comparable with the
latter component much more variable than 
the former. This state is called the {\it very-high-state} to 
distinguish it from the 
{\it high-state} which occurs later. In the high state, the soft component 
remains, but 
the hard component is very weak or totally disappears. During this state,
a ``reflare'' in the light curve (by a factor about 2) occurred about 70 
to 80 days after peak (see Ebisawa et al. 1994).
It is important to note that this same kind of reflare is present in the
light curve of BHC X-ray transients A 0620--00 and GS 2000+25, which
therefore suggests a common origin (Chen, Livio, \& Gehrels 1993).
At lower luminosity levels (about a factor of 100 from the peak value) 
Nova Muscae exhibited a {\it low-state} in which the spectrum is
very similar to that of Cyg X-1.

A unified view for the spectral states of black hole candidate sources has  
yet to emerge. Recently it has
been suggested that the spectral changes are related to  the mass flow 
rate in the accretion disk (van der Klis 1995, Nowak 1995). However, this 
was adopted as a working hypothesis and no physical explanation was provided. 
We note that an interpretation of the transition from a low to a high state as 
due to an increase of mass accretion rate is opposite to the standard accretion
disk theory which predicts thermal instabilities (and a possible transition to 
an optically thin state) only for high mass accretion rates (e.g., Piran 1978).

Accordingly, in this paper we present a new model for the interpretation of 
spectral states in black hole candidate systems in terms of accretion disk 
models characterized by an optically thick component, advection-dominated 
component, and an optically thin corona.  We suggest that the very high, 
high, low, and off states are a sequence characterized by decreasing mass 
accretion rates (see also van der Klis 1995).
A similar picture, but differing in detail, has recently been suggested by 
Narayan \& Yi (1995) and Narayan (1996).
In the next section, the model is presented in terms of 
current ideas in accretion disk theory.  The implications of such a model and 
its observational consequences are discussed in the last section. 

\section{MODEL}

Recently, Chen et al. (1995) have presented a unified description of accretion 
disks around black holes. They showed that, at a given
radius $r$, exactly four physically distinct types of accretion disks may
exist. Two of them correspond to a low viscosity in the disk, $\alpha < 
\alpha_{crit}(r)$,
and the other two correspond to a high viscosity, $\alpha > \alpha_{crit}(r)$.
These disks are further differentiated by the optical depth in the 
disk in the former case (i.e., optically thick or thin). For the high viscosity
case, the type of disk solution depends upon whether advective energy transport 
is negligible or dominant, and advection-dominated solutions exist 
for all values of the mass accretion rate. On the other hand, for a small
viscosity, advection-dominated solutions exist except for a gap 
near the Eddington rate, where no stable solution is possible and 
instability may occur. More recent developments in disk models, which have 
included the
detailed microphysics of the hot plasma, have shown that $\alpha_{crit}$ may,
in fact, exceed unity (Bj\"ornsson et al. 1996, Kusunose \& Mineshige 1996).
Therefore, we consider only the case of $\alpha < 1 \lta \alpha_{crit}$.

In Figure 1a the thermal equilibria of an accretion disk is illustrated at a 
fixed radius for the case in which the viscous stress is proportional to the 
total pressure. The solution at high column densities is optically thick and
exhibits an {\bf S}-shaped curve. The solid curve denotes thermal-viscous 
stability whereas
the dotted curve denotes instability.  The disk is stable at low mass accretion
rates where it is gas pressure dominated and at high mass accretion rates 
where it is advection-dominated.  Instability occurs when the ratio of gas
pressure to total pressure is less than 0.4.  In contrast, the solution at low 
column densities is optically thin. Here, the lower branch
is local cooling dominated (only bremsstrahlung cooling is included)
and is thermally unstable, whereas the upper
branch is advection-dominated and is thermally stable
(Narayan \& Yi 1994, Abramowicz et al. 1995a).
We identify the maximum mass accretion rate (due to the radial advection) 
at the tip of the optically thin branch as $\mdot_1$ and
the mass accretion rate at the lower turning point (due to the radiation 
pressure effects, $p_{rad}/p_{gas} \ge 3/2$) of the optically thick branch 
as $\mdot_2$. Note that at the upper turning point
(due to the radial advection again), 
$\mdot$ is usually larger than $10-50\mdote$
(see the {\bf S}-curves of Abramowicz et al. 1988 and Chen \& Taam 1993).
Hence, we consider only $\mdot$ less than that value. 
We estimate,
$$\mdot_1/\mdote=2.0 \times 10^{3} f (r/r_g)^{-1/2} \alpha^{2}, \eqno(1)$$
$$\mdot_2/\mdote=4.0 \times 10^{-3} f^{-1} (r/r_g)^{21/16} (\alpha
M/\msun)^{-1/8}. \eqno(2)$$
Here ${\dot M}_E = 4\pi GM/(c\kappa_{es})$ is defined as the Eddington rate, 
where $\kappa_{es}$ is the electron scattering opacity taken to be equal to 
0.34 and $f=1-\ell_*/\ell$ is considered for the
inner boundary effect, where $\ell$ and $\ell_*$ are the specific angular 
momenta at radius $r$ and the inner edge $r_*$ respectively.
The specific angular momentum is Keplerian and is calculated under the
pseudo-Newtonian potential of Paczy\'nski \& Wiita (1980).
Formula (1) is calculated as that in Abramowicz et al. (1995a)
where only the bremsstrahlung is included for the local cooling.

The variations of $\mdot_1$ and $\mdot_2$ with respect to the radius of
the disk are shown in Figure 1b.  It can be seen that $\mdot_1$ 
decreases slowly with increasing radius in the outer part of the disk. 
This trend reflects the fact
that local cooling processes become more efficient relative to the heating rate
at larger radii. 
A similar formula for $\mdot_1(r)$ has been calculated by 
Narayan \& Yi (1995) under the self-similar assumption.
It has the same scaling with $\alpha$
but the coefficient is smaller by a factor of approximately 10.
In their calculation, 
the detailed radiative cooling processes such as synchrotron 
and Componization are included.
Since these processes become more important for smaller radii, $\mdot_1$
decreases as $r$ decreases. In their case,
$\mdot_1/\mdote 
\approx 10-20\alpha^2$ for $r \lta 1000r_g$, while our formula (1)
gives $\mdot_1/\mdote \approx 100\alpha^2$ for $r \approx 1000r_g$.
More recent calculations have shown that, with similar detailed microphysics,
the self-similar solution gives a smaller $\mdot_1$ than that calculated
under the Keplerian disk assumption. For example, for $M=10\msun$,
$\alpha=0.1$, and $r=30r_g$,  
Bj\"ornsson et al. (1996) obtained $\mdot_1/\mdote \approx 0.4$ 
(see also Kusunose \& Mineshige 1996) while Narayan \& Yi (1995) 
gave $\mdot_1/\mdote \approx 0.035$ (see their Figure 1 and scale $\mdot_1$ to
the same $\alpha$ and $\mdote$).
There are probably two factors contributing to this difference.
One is the self-similar assumption which underestimates the value of the 
angular velocity (see global solutions of Chen, Abramowicz, \& Lasota 1996 and
Narayan, Kato, \& Honma 1996a) and thus results in a smaller heating rate.
A smaller heating rate shifts the advection dominated thermal equilibrium 
curve lower and shifts the optically thin local cooling dominated  
thermal equilibrium curve higher, therefore $\mdot_{1}$ becomes smaller. 
Note that the Keplerian approximation overestimates the angular velocity 
and, hence, the heating rate.  Therefore, $\mdot_1$ lies within the range
determined by these approximations.  
The second factor is related to the inner boundary 
condition which cannot be included in the self-similar solution.
An accurate $\mdot_1$ can only be obtained by constructing global
solutions of the disk including detailed radiative microphysics.
For the purpose here, considering both effects of the radiative microphysics
and the angular velocity,
we may assume $\mdot_1$ as a very weak function of $r$, defined as
$\mdot_{c1}$ for $r \lta 1000r_g$:
$$\mdot_{c1}/\mdote \approx 50 \alpha^{2}. \eqno(3)$$
Note that even a smaller $\mdot_{c1}$ as that of Narayan \& Yi (1995) will 
not effect our conclusion later, since a slight increase of $\alpha$ will 
easily compensate for it.

The increase of $\mdot_2$ with radius in the outer part of the
disk is attributable to the tendency that radiation pressure becomes important 
at larger radii only at higher mass accretion rates.   Figure 1b clearly shows
that $\mdot_2$ has a minimum of $\mdot_{c2}$ at radius of about 
$r_c \approx 8-9\, r_g$, where $r_g = 2GM/c^2$:
$$\mdot_{c2}/\mdote=0.2 (\alpha M/\msun)^{-1/8}. \eqno(4)$$
It is clear from Figure~1b that no steady solutions exist for $\mdot \gta
\mdot_{c1}$. That is, in this mass accretion rate range the region 
characterized by $r<r_2$ is non steady (see Figure 1b).
We note that this region could be 
extensive, $r_2 \sim 100 r_g$, if $\mdot$ is large.
It can be calculated that for $\mdot \approx \mdote$ and $\alpha \gta 0.1$, 
the effective optical depth of the disk in regions $r \lta 100 r_g$ is less
than unity. We suggest that this region is hot and optically
thin and is the seat of the chaotic hard X-ray variability. 
Outflow is highly likely especially if $\mdot > \mdote$. 
The region exterior to $r_2$ is optically thick which produces the soft X-rays.
The hard and soft X-ray luminosities may be comparable depending on the 
location of $r_2$. This state is identified with the {\it very high-state}.

At lower accretion rates, $\mdot_{c1}>\mdot>\mdot_{c2}$, region $r<r_2$ is 
described by an advection-dominated 
flow which is hot, optically thin and stable (see also Narayan \& Yi 1995).
It is seen that, as $\mdot$ decreases to near the bottom of the line $r_2$,
the optically thick gas pressure dominated disk solution starts to be 
available in the region to the left of the line $r_2$ inside a few $r_g$.  
The global solution in this small region will however remain optically thin
because its "outer" boundary condition is optically thin and
advection dominated at this stage.
In addition, since the flow is transonic and 
the radial drifting time is almost a free-fall time scale,
the once hot advection dominated flow does not have time to settle to 
a cool state there.
Since most of the gravitational binding energy is advected
into the black hole, the disk produces very little hard X-rays.  On the other
hand, the region exterior to $r_2$ is optically thick and the soft 
X-ray emission
dominates the total luminosity. This disk configuration describes the  system 
in its {\it high-state}.  Note that, as the mass accretion rate decreases, the
ratio of the hard X-rays to the soft X-rays decreases since the spatial extent
of the advection-dominated region decreases while that of the optically thick 
region increases. 

As the mass accretion rate declines further, $\mdot_{c2}>\mdot>\mdot_{c3}$, 
the entire disk becomes optically thick
and the spectrum is similar to that described above for the high state. 
This structural change, however, may result in an
increase of luminosity, corresponding to a reflare, due to the sudden
decrease of the inner radius of the disk, $r_{in}$, 
from $r_2=r_c \approx 8-9 r_g$ to
$r_2 \sim 3 r_g$ (since now, the whole disk can be optically thick).
Since the disk luminosity is approximately inversely proportional to the 
inner disk radius, its shift results in a luminosity increase by a factor of 
$\sim 2-3$.

A transition to a new stage occurs for $\mdot_{c3}>\mdot>\mdot_{c4}$, where 
the system is in a corona-disk
configuration. The formation of corona above the disk may be due to an 
evaporation mechanism facilitated by the electron conduction process as 
envisaged by Meyer \& Meyer-Hofmeister (1994). 
The corona is hot and optically thin
and is responsible for the production of the hard X-rays. The underlying 
disk is optically thick. This state corresponds to the {\it low-state}. 
In this corona-disk model, the soft X-rays (produced by the
viscous dissipation in the underlying cold disk as well as by the 
reprocessing of the hard X-rays from the above corona) are also Comptonized
in the corona. Therefore the overall energy spectrum can be 
described as a power-law (Haardt et al. 1993). 
The soft blackbody component, however, has a temperature of only 
$T_{bb} \lta 0.1$ keV because of the low mass accretion rate. This
component, however, is difficult to detect since the spectrum peaks in
the wavelength region where interstellar absorption is high. 

At the lowest rates of mass accretion, $\mdot<\mdot_{c4}$, the electron 
conduction process may lead to the formation of a totally optically thin disk. 
In this case, the flow is advection-dominated, thereby producing
very little luminosity corresponding to the {\it off-state}.  In particular, 
Narayan, McClintock, \& Yi (1996b) have applied the 
advection-dominated accretion flows to the off-state of A 0620--00
(see also Lasota, Narayan, \& Yi 1996).
We note that Narayan (1996) has applied the advection-dominated accretion 
flows to the low-state of BHCs, which however require a high viscosity 
parameter ($\alpha \sim 1$) to produce the observed X-ray luminosity. 
In addition, the low-state
has also been modeled by a small inner advection-dominated flow plus 
an outer disk-corona structure  by Abramowicz, Chen, \& Taam (1995b).
The model, presented here, may reconcile these differences depending on the 
values of $\mdot_{c3}$ and $\mdot_{c4}$.

The critical values of $\mdot_{c3}$ and $\mdot_{c4}$ are difficult to estimate,
but further work along the directions advocated by  
Meyer \& Meyer-Hofmeister (1994) on CV's may be fruitful. For example, 
$\mdot_{c4}$ may be
determined by the balance between the evaporation rate into the corona 
and the mass accretion 
rate of the disk. 

A schematic diagram illustrating the above description for the systematic 
variation of the disk configuration with declining mass accretion rates is 
shown in Figure 2. 

\section{DISCUSSION}

An interpretive framework for the spectral states of black hole candidate 
sources has been presented in terms of the mass accretion rate only.  The 
accretion disk models upon which the framework is based require
a moderately large $\alpha$. Since
$\mdot_{c1} \propto \alpha^{2}$, both very small and very large $\alpha$
will give an extremely small or large $\mdot_{c1}$ respectively, which
is largely excluded by the interpretation of observational data (see below). 
Considering the uncertainties involved with $\mdot_{c1}$, 
we estimate $\alpha \sim 0.1-0.3$, which 
gives an $\mdot_{c1}$ corresponding to a luminosity 
near the Eddington limit. 
This is probably the case for Nova Muscae.
This value of $\alpha$ is
consistent with the disk instability model which produces the outburst
and fits the exponential decay of the light curve
with an e-folding time of about 30-40 days
(Cannizzo, Chen, \& Livio 1995, and references within).

It is seen that the critical mass accretion rate, $\mdot_{c2}$, below which 
the disk can be stable and optically thick depends on the disk parameter very 
weakly.  Since it corresponds to the luminosity level 
where the ``reflare'' occurs, it provides
an estimate of the absolute value for the luminosity for a given accretion 
efficiency.  If the distance to the source and the inclination angle of the 
binary system are known, the observed luminosity can be
used to provide an estimate of the mass of the black hole.
Independent of the distance to the source, 
$\mdot_{c1}/\mdot_{c2}$ can be determined from the ratio of the
peak flux during the high  state 
to the flux just before the reflare. The observations of Nova Muscae 
(Ebisawa et al. 1994) and GS 2000+25 (Tsunemi et al. 1989) indicate
that this ratio is about $\sim 7-10$, 
which is consistent with $\alpha \approx 0.1-0.3$.

In the present model, the ``reflare'' occurs as a result of the inward shift 
of the inner edge of the optically thick disk. 
Thus, the "reflare" is unique to a black hole source, with significant  
emission in the soft X-ray band expected.
In fact, this model predicts a decline in hard X-rays during the reflare
(see, for example, Ebisawa et al. 1994). This phenomenon cannot be explained
by other models (see Wheeler et al. 1996 for a review). 
It also predicts that, if there is also an 
optical/UV ``reflare'', it would result from the reprocessing of X-rays and, 
thus, it should be delayed in this model.
This is in contrast to the irradiation (of the secondary star) model
suggested by Chen et al. (1993) and the irradiation of the outer disk as 
envisioned by Kim et al. (1995) in which, the optical/UV ``reflare'' is
expected to arise prior to the X-ray reflare (see also Mineshige 1994).

Our model does not predict a very high state if the accretion rate is less than
$\mdot_{c1}$. In this case, the ``reflare'' should occur earlier provided that 
the mass accretion rate is greater than $\mdot_{c2}$. This may  
be the case in A 0620--00 in which the "reflare" occured after 50--60 days 
after peak luminosity (see the light curve in Kaluzienski et al. 1977 and
Tanaka \& Lewin 1995).

Our model also implies that, if a source has no high-state (i.e., 
stays hard), then there should be no reflare. This appears to be the case in
GS2023+33 (V404 Cyg: Kitamoto et al. 1989, Miyamoto et al. 1995)
and in GRO J0422+32 (Nova Persei: Sunyaev et al. 1993,
Callanan et al. 1995, Vikhlinin et al. 1995).
Note that the big-bump of hard X-rays in J0422+32 which occurred more 
than 100 days after peak (Callanan et al. 1995) may
have the same origin as the second larger reflare observed in
A 0620-00 and Nova Muscae,  which we believe is different from the
``reflare'' we have studied here (see Chen at al. 1993).

We note here that although the efficiency of producing luminosity is
very different, the energy spectra from the optically thin hot medium,
of the inner region nonsteady flow in the very high-state, the corona
in the low-state, and the advection dominated flow in the off-state, 
are rather similar since the radiative cooling mechanism is similar
(for a spectral calculation, see Narayan et al. 1996b).
The contribution of soft X-rays from the outer optically thick disk
in the very high-state is easily observed since the mass accretion rate
there is high and the typical blackbody temperature 
is around 1keV. On the other hand, in the low state,
the underlying cold disk inside the corona has a very low
blackbody temperature, $\lta 0.1$ keV, and therefore, its detection
is difficult. Interesting, Baluci\'nska-Church et al. (1995) recently observed
a soft X-ray ($T_{bb} = 0.13\pm 0.02$ keV) excess in Cyg X-1 and they
identified it as the disk emission. The luminosity of this soft excess
is about $4.7 \times 10^{36}$ erg\,s$^{-1}$ assuming a distance of the source
as 2.5\,kps.

Narayan (1996) modeled the low-state as an advection-dominated accretion
flow. In the black hole X-ray transient case, then, the luminosity
should drop significantly from the high state (local cooling dominated)
to the low state (advection dominated), independent of the viscosity
parameter. This has however not been observed and is 
unlike the off-state where the X-ray luminosity and the mass accretion rate
inferred from other observations indicate a very low luminosity
efficiency implying an advection dominated disk (Narayan et al. 1996b).
There are other observational tests to distinguish the advection
dominated model from the 'sandwiched' 
corona-disk model for the low state. For example,
in the latter case, the spectrum should consist of a reprocessed (in the
corona above) hard X-ray reflection (off the disk) component. 
Haardt et al. (1996) have constructed a corona-disk model to explain 
this type of spectral component from Cyg X-1 (Gierli\'nski et al. 1995). 
Theoretically, time-dependent simulation of both types of models
could determine whether or not the chaotic variability of the low-state can
be reproduced within each model.

In conclusion, the unified description of accretion disks around black holes 
have 
appealing characteristics which can be applied to the spectral states of black 
hole candidate systems.  In the proposed model the spectral states are governed
by the mass accretion rate alone.  This picture, however, is incomplete since
the evidence from a combination of observational results from a number of BHC 
sources reveal that the luminosity level at which the transition from a low 
state to a high state on the ascending portion of the outburst differs from the 
transition from the high state to the low state during the descending portion 
by a factor of about 100 (see Miyamoto et al. 1995). A possible explanation 
for this hysteretic behavior is that the transition from the low to high 
state is physically distinct from the transition from the high to low state. 
For example, the spectral state during the rising portion of the outburst may 
correspond to an optically thin disk in its advection-dominated state (i.e. 
the solution corresponding to low column densities in Fig. 1a).  The disk may 
remain advection-dominated until a transition occurs 
to the high soft state at accretion rates near $\mdot_1(r)$.
On the other hand, the transition from the high state to 
the low state on 
the decay phase of the outburst reflects the operation of the evaporation 
mechanism in the corona.  Thus, the transition from optically thin to 
optically thick 
and vice versa are not necessarily symmetric and differences in the 
luminosity level at which such transitions occur may be expected to be 
different.  To confirm the framework outlined in this paper and to quantify 
the interpretation for such transitions, models for accretion disks
based on a global solutions rather than the local solutions must be 
constructed.  The implications of such solutions on the 
hysteretic behavior inferred in observed sources remain for future
investigations.

\acknowledgements

\n X.C. thanks Jean Swank for hospitality and discussions while he was
a visitor at GSFC/NASA during March 1995. 
We thank Jean-Pierre Lasota and Ramesh Narayan for their comments on an 
earlier version of the manuscript.
This research was supported in part by NASA under grant NAGW-2526.

\newpage
\begin{figure}
\caption{(a) The thermal equilibria of an accretion disk in the mass accretion
rate-column density plane. Two critical mass accretion 
rates are denoted by $\mdot_1$ and $\mdot_2$, corresponding to the maximum
rate for an optically thin disk and the first turning point of the 
optically thick disk respectively.~
(b) The radial variations of $\mdot_1$ and $\mdot_2$ (solid lines). 
The dotted line is for $\mdot_{c1}=2$ from equation (3).
Note that $\mdot_2$ has a minimum of $\mdot_{c2} \approx 0.23$ 
at about $r_c \approx 8-9 r_g$.}
\end{figure}

\begin{figure}
\caption{A schematic description of the disk configurations and the spectral 
states of black hole candidates (see text).}
\end{figure}
\end{document}